\documentclass[prd, twocolumn, nofootinbib]{revtex4}

\usepackage{epsfig} 
\usepackage{amsmath}

\newcommand{\beq}{\begin{equation}}
\newcommand{\eeq}{\end{equation}}
\newcommand{\beqa}{\begin{eqnarray}}
\newcommand{\eeqa}{\end{eqnarray}}

\newcommand{\ls}{\mathrel{\raise0.27ex\hbox{$<$}\kern-0.70em \lower0.71ex\hbox{{
$\scriptstyle \sim$}}}}

\begin{document} 

\title{Like vs.\ Like: Strategy and Improvements in Supernova Cosmology 
Systematics} 
\author{Eric V.\ Linder} %
\affiliation{Berkeley Lab \& University 
of California, Berkeley, CA 94720, USA} 
\date{\today}

\begin{abstract} 
Control of systematic uncertainties in the use of Type Ia supernovae 
as standardized distance indicators can be achieved through contrasting 
subsets of observationally-characterized, like supernovae.  Essentially, 
like supernovae at different redshifts reveal the cosmology, and 
differing supernovae at the same redshift reveal systematics, including 
evolution not already corrected for by the standardization.  Here we 
examine the strategy for use of empirically defined subsets to minimize 
the cosmological parameter risk, the quadratic sum of the parameter 
uncertainty and systematic bias.  We investigate the optimal recognition 
of subsets within the sample and discuss some issues of observational 
requirements on accurately measuring subset properties. 
Neglecting like vs.\ like comparison (i.e.\ creating only a single 
Hubble diagram) can cause cosmological constraints on dark energy to be 
biased by $1\sigma$ or degraded by a factor 1.6 for a total drift of 
0.02 mag.  Recognition of subsets at the 0.016 mag level (relative 
differences) erases bias and reduces the degradation to 2\%. 
\end{abstract} 

\maketitle

\section{Introduction \label{sec:intro}}

Distance-redshift measurements of Type Ia supernovae (SN) provide direct 
mapping of the cosmic expansion history.  The peak brightness of most SN 
have tighter dispersion than any other cosmological object and this can 
be standardized with a simple light curve amplitude-width relation, first 
established by \cite{phillips} in the early 1990s. 
This allows a SN to be calibrated 
to 0.15 mag or about 7\% in distance, and provided the technique to discover 
the accelerated cosmic expansion in the late 1990s \cite{perl99,riess98}. 
See \cite{leibundgut} for a review as of 2001.  
For revealing the nature of the physics causing the acceleration, generically 
called dark energy, SN have continued to play a central role (e.g.\ 
\cite{knop03,riess04,astier05,riess06,essence,davis07,komatsu08,kowalski,
rubin}).  

The limits to the standardization for SN are not known; a second parameter 
to further reduce the intrinsic dispersion is actively sought among the 
SN observables (see, for example, \cite{aspen}), and more detailed 
measurements in spectroscopy and a wider 
range of wavelength bands may turn up new observables and correlations. 
The uncertainty on cosmological parameters improves as the intrinsic 
scatter decreases, both more rapidly than linearly as the reduced 
dispersion further improves color and dust corrections, and less rapidly 
as measurement uncertainties remain. 

Reduction in scatter can also be achieved by characterizing each supernova 
with a detailed array of measurements, expecting that supernovae with 
identical observed properties must also have identical intrinsic luminosities. 
These empirical observations can define subsets of SN.  Note that the converse 
does not necessarily hold -- SN that differ in some property, e.g.\ position 
in the host galaxy or its metallicity, may not diverge in luminosity 
(see \cite{howell08100031} for one recent study).  This 
was referred to as the mapping of subsets (empirical differences) to 
subclasses (intrinsic luminosity differences) in \cite{snlow}. 

Mere differences in luminosities are not sufficient to affect cosmological 
parameter estimation, since they will be absorbed into the ``nuisance'' fit 
parameter for the intrinsic luminosity (which will be impacted).  A further 
ingredient must be present: population drift, or evolution of the relative 
fraction of each subclass with redshift.  Note that SN are not per se aware 
of the Hubble expansion: the explosions and radiation transport take place 
on scales $10^{-13}$ times smaller than the Hubble length.  So SN should not 
evolve in a cosmic sense; rather they may be affected by their immediate 
environment and progenitor conditions.  Since the full diversity of 
environments from higher redshifts also exists at low redshifts (e.g.\ stars 
and galaxies continue to form today), only the proportion of different 
environments changes, population drift is a more accurate description of 
possible changes in SN luminosity.  

It is important to note here that while 
subsets of SN have been recognized, and the proportion of some subsets 
has been seen to change with redshift, {\it current data show no definite 
indication that SN luminosity evolves\/} -- other than is automatically 
corrected for in using a standard single parameter light curve 
amplitude-width relation.  That is, we know of no subsets that are 
subclasses. 

This article looks to the future when suites of observations on large samples 
of SN, more detailed measurements than we have on any individual SN today, 
{\it may\/} show that indeed some subset, defined through those 
observational characteristics, is a subclass having a different luminosity.  
The basic method of comparing subsets of like SN -- likes vs.\ likes, or SN 
demographics -- was explained clearly in \cite{coping}, and we follow this 
approach while extending it to calculating detailed effects on cosmological 
parameter determination. 

Systematics is emphatically the name of the game in accurate science. 
Understanding the level of control is essential: without an intrinsic 
floor, SN are only limited by cosmic variance (from the number of SN 
within a Hubble volume) to 0.003\% in distance precision.  And of course 
a biased answer can be worse than an imprecise one.  

For the reader wanting a quick conclusion, see Fig.~\ref{fig:riskfomsimple}. 
In \S\ref{sec:pdf} we establish the formalism of subclass luminosity functions 
and calculate the effects of population drift on the mean and variance of the 
full sample luminosity function.  Using this in \S\ref{sec:cos}, we identify 
three distinct impacts on cosmology determination, and show that the bias 
is a major effect.  In \S\ref{sec:risk} we examine the interplay of bias 
and uncertainty as we investigate strategies for controlling systematics, 
such as adding fit parameters for observationally recognized subclasses.  
We address aspects of the observational requirements for 
identifying subclasses in \S\ref{sec:obs}, and conclude in \S\ref{sec:concl}. 

\section{Subclasses, Population Drift, and Magnitude 
Evolution \label{sec:pdf}} 

We begin by considering an observed sample of SN to be composed of a set 
of subsamples 
that we may or may not distinguish.  The intrinsic luminosity, or magnitude, 
distribution of the overall SN population at some redshift is a sum over 
all the individual subset luminosity functions.  That is 
\beq 
\Phi(L,z)=\delta(L-\sum f_j(z) L_j)\ \sum f_i(z)\phi_i(L) \label{eq:lumpdf} 
\eeq 
where $\phi_i$ is an individual subset luminosity function, $L_i$ the 
mean luminosity of that subset, and $f_i(z)$ the fraction of the total 
population sample that subset represents at redshift $z$.  

Note again 
that $L_i$ is the mean luminosity: we are not imposing that the subset 
is a subclass with standard luminosity\footnote{Of 
course more tightly defined 
subsets should have smaller dispersion about the mean, and poorly 
characterized subsets, or those defined through variables unrelated to 
the luminosity, may have larger dispersion such that the difference 
between subsets' luminosities is smeared out, but this does not affect 
the formalism.  See \S\ref{sec:obs} where we return to discussion of 
these points.}, 
only that the {\it mean\/} luminosity is independent of redshift.  
This still places a strong burden on excellence of observations 
and requires something in between an empirically defined subset (since 
we need some knowledge of the luminosity behavior, i.e.\ the subset of 
SN discovered on a Tuesday is insufficient) and a subclass.  We 
discuss this challenge further, and how to handle deviations, in 
\S\ref{sec:defsub}.  For now we continue to call it a subset, and 
effectively each subset's $L_i$ represents a different absolute magnitude 
${\mathcal M}_i$. 

Given Eq.~(\ref{eq:lumpdf}) for the probability distribution function we 
can calculate whichever moments of the total luminosity distribution 
desired, in terms of moments of the individual subset luminosities, 
without requiring any assumption of, say, a Gaussian form.  
The drift of the mean luminosity of the sample, relative to 
the value at some redshift $z_0$, is 
\beq 
\langle L(z)-L(z_0)\rangle = \sum \delta L_i(z_\star)\,[f_i(z)-f_i(z_0)], 
\label{eq:dL} 
\eeq 
where we define the offset of each subset mean luminosity as 
\beq 
\delta L_i(z_\star)=L_i-L(z_\star).
\eeq 
We are free to evaluate the subset offset relative to some other redshift 
$z_\star$, though generally we will take $z_0=z_\star=0$. 

The result for the variance of the total sample luminosity is 
\beqa 
\sigma_L^2(z)&\!&=\langle L^2(z)\rangle-\langle L(z)\rangle^2 \\ 
&\!&=\sum f_i(z)\,\sigma_i^2 \label{eq:sigL} \\ 
&\!&\qquad +\ \sum f_i(z)\,\delta L_i^2(z_\star)- 
\left[\sum f_i(z)\,\delta L_i(z_\star)\right]^2\,. \nonumber 
\eeqa 
The first term is a subset-weighted dispersion, where $\sigma_i^2$ is 
the luminosity variance of subset $i$, and the final two terms are 
contributions from the offset of the mean subset luminosity relative to 
the mean sample luminosity.  If the offsets $\delta L_i$ are zero (if 
they are equal, they must be zero by the delta function in 
Eq.~\ref{eq:lumpdf}), then these bias terms vanish.

\section{Effects of the Magnitude Distribution on Supernova 
Cosmology \label{sec:cos}} 

Recognizing subclasses of SN can have three effects on the calculation 
of cosmological parameter uncertainties: it might 1) reduce the dispersion 
of the sample used in the Hubble diagram, 2) reduce the residual systematic 
error, 3) reduce cosmology parameter bias if analyzed in the proper way. 

For the first effect, let us first consider the influence of the offset 
terms in Eq.~(\ref{eq:sigL}).  The following argument indicates they likely 
do not have a substantial impact.  Consider two subsets, offset in 
absolute magnitude by $\delta m_{12}$.  (For the remainder 
of the article we phrase the analysis in terms of magnitudes rather 
than luminosities; for small differences in subset luminosities one 
can make a direct substitution in the formulas.)  Then 
\beq 
\sigma_m^2(z)=\sigma_2^2+f_1\,(\sigma_1^2-\sigma_2^2)+ 
\delta m_{12}^2\,f_1(1-f_1), 
\eeq 
where $f_1$ is fraction of the population in subset 1 (and $1-f_1$ is in 
subset 2).  Since the maximum of $f_1(1-f_1)$ is 1/4, and the magnitude 
offset should be (much) less than the dispersion, the last term is 
unlikely to be important.  For $\sigma_1\sim\sigma_2$, we 
simply have that the dispersion of the sample is nearly the dispersion 
of the subsets. 

Next within effect 1 we consider when the dispersion in a subsample is 
reduced.  While 
this has a mild effect on the variance of the full sample, we can imagine 
a Hubble diagram formed only from the subsample (as suggested for example 
for elliptical galaxy hosted SN, though this reduces the external 
systematic of dust extinction not the internal luminosity variation).  
This will have fewer data points, decreasing the cosmological leverage 
in opposition to the lesser dispersion.  
In the statistical error regime, the 
error is effectively $\sigma_i/\sqrt{N_i}\sim\sigma_i/f_i^{1/2}$, so the 
subset must account for a sizeable fraction of the population over a 
wide range of redshifts in order for this subdiagram to improve in 
precision over the full Hubble 
diagram.  For example, if $\sigma_{\rm full}=0.15$ and $\sigma_1$=0.1, 
one requires $f_1>0.44$.  Data from ongoing large surveys, such as 
the Supernova Legacy Survey (SNLS \cite{snls}), the Nearby Supernova 
Factory (SNf \cite{snf}), and the CfA Supernova Archive \cite{matheson}, 
that characterize SN properties in detail may lead to such 
reduced-dispersion subsets.  However, future surveys are likely to be 
fundamentally limited by systematics.  In the systematic error 
regime, while one has supernovae to spare and could use only those from 
the reduced-dispersion population, the systematics dominate over the 
statistical dispersion and the subset Hubble diagram does not help, unless 
effect 2 enters, reducing systematics. 

As we obtain more incisive measurements of the supernova sample, 
characterizing each SN in more detail, we can potentially reduce the 
systematic uncertainties.  Recall that systematics refers to the 
uncertainties remaining after correction procedures have been applied, so 
the more information about a SN, the better chance for corrections to work 
to a deeper level.  In the systematics dominated regime, an improvement 
by a factor of two in systematics leads to a factor of two tighter 
constraints on cosmological parameters.  This is a strong reason 
for gathering a large suite of measurements to recognize subsets.  However, 
not all systematics in an experiment arise from source properties -- 
instrumental errors such as from filters and calibration also enter.  
The effect of subset recognition on residual systematics requires detailed, 
experiment-specific simulations for quantitative answers.  Thus, although 
such data holds considerable promise for improving supernova probes, 
in this paper we concentrate on the third effect, cosmological parameter 
bias and degradation, treated in the next section.

\section{Minimizing Systematics Impact on Cosmology 
Determination \label{sec:risk}} 

Without recognition of subclasses, population drift among them will 
appear as evolution in the mean absolute magnitude of the sample.  
Again, the differential population demographics is key -- mere constant 
differences between subclasses are absorbed completely into the fit 
parameter for the absolute magnitude, ${\mathcal M}$, and do not 
impact the cosmological parameters.  The bias induced by the magnitude 
evolution on the cosmological parameters can be evaluated at the same 
time as the parameter estimation uncertainties with standard Fisher, 
or information, matrix techniques.  

Specifically, the bias on parameter $p_i$ is 
\beq 
\delta p_i=(F^{-1})_{ij}\sum_k\frac{\partial m_k}{\partial p_j} 
\frac{1}{\sigma_k^2}\Delta m_k, \label{eq:bias} 
\eeq 
where $m_k$ is a supernova magnitude, $\Delta m_k$ the offset due to 
the effective evolution, and 
$F$ is the Fisher matrix over parameters $p$.  For simplicity, we 
write this for a diagonal error matrix with entries $\sigma_k$; see 
\cite{linbias} for the generalization. 

As an example of the importance of accounting for bias, note that for 
two populations differing in absolute magnitude by 0.02, with one 
dominating at low redshift and  the other at high redshift, the 
cosmological parameter bias can amount to greater than one 
statistical sigma. 

Only two ingredients are required for determining the impact on 
cosmology: the mean absolute magnitude of each population, $M_i$, and 
the population fractions or demographics $f_i(z)$, which combine 
together to form the offset $\Delta m$ through Eq.~(\ref{eq:dL}), 
\beq 
\Delta m(z)=\sum \Delta M_i\,[f_i(z)-f_i(0)]\,,  
\eeq 
where $\Delta M=-2.5\log\delta L_i(0)$.  
The first ingredient is of course not known from observations, while 
one could measure the populations $f_i$ from the data itself (we 
discuss this further in \S\ref{sec:obs}).  We consider several 
different models for each and examine the range of cosmology impacts. 

To remove a bias induced by different $M_i$, one could introduce 
additional fit parameters for them (or equivalently for 
${\mathcal M}_i=M_i-5\log h$, where $h$ is the dimensionless Hubble 
constant).  This of course only applies to those subsets that 
are {\it recognized\/}, i.e.\ empirically distinguished  
by the values of a certain set of measurements (for example, high line 
velocity, elliptical galaxy host, strong ultraviolet flux, etc.).  The 
more subsets recognized, and fit parameters introduced, the less bias 
in the cosmology determination, but the more uncertainty in the estimation 
of the cosmology parameters due to the larger parameter space. 

Explicitly, if there are $N$ subsets and we have the observational acuity 
to recognize $R$ of them, then we can fit for $\mathcal{M}$ (representing 
the gross subsample of unrecognized subsets) and 
$\mathcal{M}_1$,\dots $\mathcal{M}_R$ and suffer a cosmology bias 
\beq 
\Delta m(z)=\sum_{i=R+1}^{N} \Delta{\mathcal M}_i\, 
[f_i(z)-f_i(0)] \label{eq:unrecog}
\eeq 
due to the $N-R$ unrecognized subsets.  The question then is simply which 
wins out: improved precision from fitting for fewer parameters, or improved 
accuracy from reducing bias.  That is, what is the optimum value for $R$ 
(given the properties $\mathcal{M}_i$, $f_i(z)$ of the subsets). 

To take into account both the dispersion and bias in parameter estimation, 
a standard statistical tool is the risk \cite{kendall}, the square root 
of the quadratic sum of the two terms, i.e.\ 
\beq 
{\rm Risk}(p)=\sqrt{\sigma_p^2+\delta p^2}\,. \label{eq:risk} 
\eeq 
We analyze the risk as a function of magnitude offsets, population 
model, and subsets recognized, seeking the optimal strategy for supernova 
cosmology -- is it better to have a single Hubble diagram of all 
supernovae, which will have tight but biased parameter constraints, or to 
divide the sample into the maximum number of recognized subsets, giving 
looser but less biased cosmology determination. 

\begin{figure}[!htb]
\begin{center}
\psfig{file=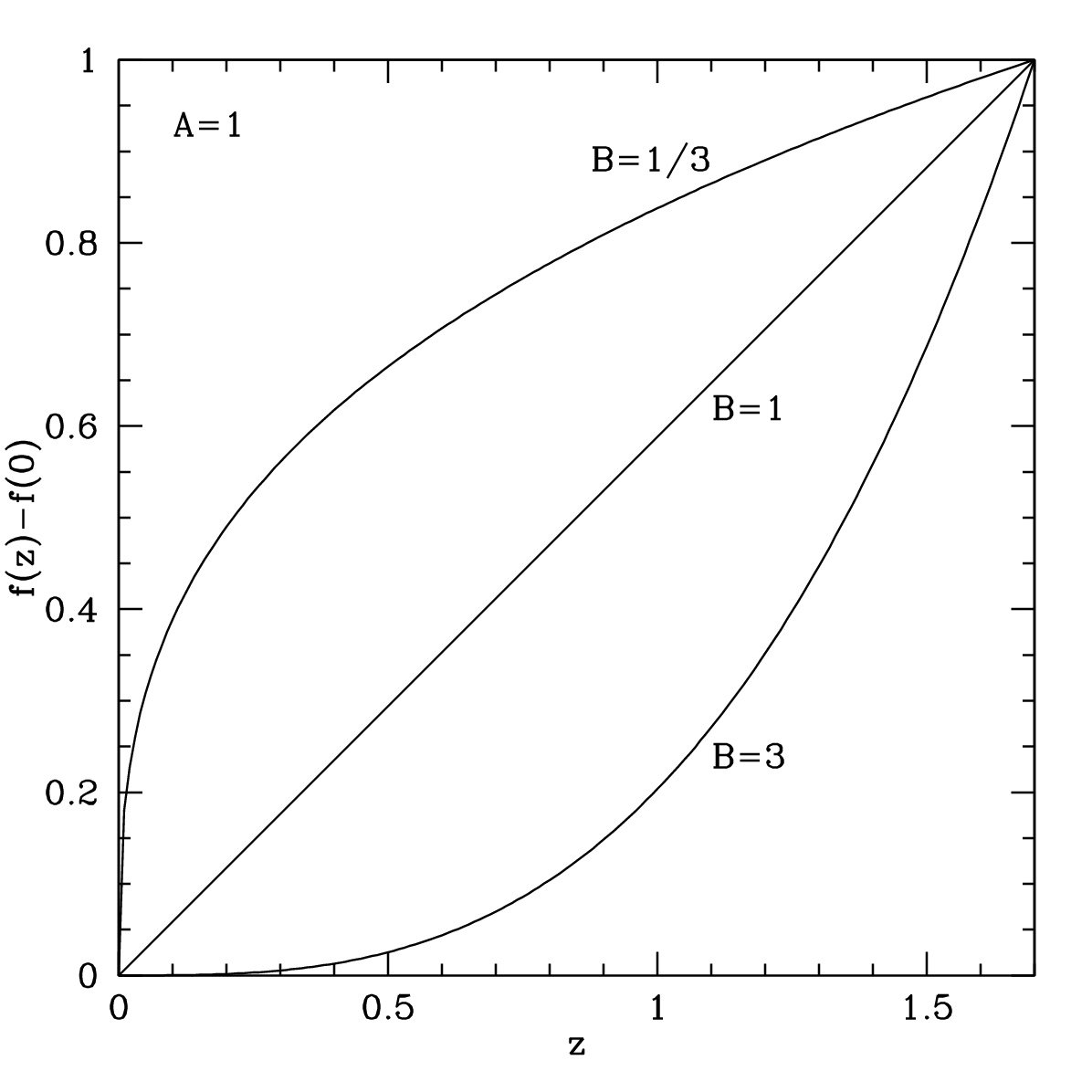,width=3.4in}
\caption{The population drift model is designed to cover a range of 
behaviors from rapid evolution at low redshift ($B<1$), to linear 
evolution ($B=1$), to rapid evolution at high redshift ($B>1$).  The 
coefficient $A$ sets the amplitude of the drift. 
}
\label{fig:fz}
\end{center}
\end{figure}

For the population model we adopt the form 
\beq 
f_i(z)=f_i(0)+A_i\,(z/1.7)^{B_i}\,. \label{eq:fz}
\eeq 
(See \S\ref{sec:defsub} and the Appendix for generalizations.)  
This is subject to the constraint that the populations sum up to the total 
sample, $\sum f_i(z)=1$ for all $z$, which is easiest to implement if 
$B_i=B$.  We consider $B=1/3$, 1, 3 to cover a range of behaviors.  
Figure~\ref{fig:fz} illustrates these population evolutions, giving 
respectively a high rate of change at low redshift, even weighting, or 
a high rate at 
high redshift.  While one could consider scale factor or cosmic time as 
the independent variable, this is not qualitatively different from 
changing the value of $B$.  Also, as a concrete example note that the 
population drift in the mean stretch parameter seen by \cite{howells} 
follows a linear redshift dependence ($B=1$) from $z=0.03-1.12$.

For the absolute magnitudes of the individual subsets we take them to 
differ from the mean absolute magnitude by $\pm X$, $\pm 2X$, 
considering four subsets.  A constant shift $\bar{\mathcal M}$ in the 
magnitudes is simply absorbed into the absolute magnitude nuisance 
parameter -- if all subsets are accounted for.  The mean absolute 
magnitude {\it is\/} relevant when only some subsets are recognized, 
since then the sum of those populations can be redshift dependent 
(i.e.\ not unity, or zero).  Explicitly, 
\beqa 
\sum{}'\,(\Delta{\mathcal M}_i+\bar{\mathcal M})\!\!&\!&\!\![f_i(z)-f_i(0)] \\ 
&\!&=\Delta m(z)+ \bar{\mathcal M}\sum{}' [f_i(z)-f_i(0)], \nonumber 
\eeqa  
where a prime denotes the sum runs over unrecognized subsets.  But 
the last term is zero only when $\sum' f_i(z)=1$ for all $z$, i.e.\ 
all subsets are included in the sum (all unrecognized), or the sum is 
trivially zero (all subsets recognized). 

The cosmology bias will scale with the subset magnitude offsets so we 
can express the results as a function of the effective magnitude 
evolution in the full sample.  That is, we can phrase the offset 
amplitude $X$ in terms of $\Delta m(z=1.7)$, say.  In the specific 
example treated below, we take $\{f_i(z=0)\}=\{1/4,1/4,1/4,1/4\}$ and 
$\{f_i(z=1.7)\}=\{1/2,3/8,1/8,0\}$, with the population drift rate 
determined by the value adopted for $B$, as in Eq.~(\ref{eq:fz}). 
In this case, $\Delta m(z)=(5X/4)(z/1.7)^B$. 

To analyze the cosmological impact, we must take into account 
both the cosmology parameter estimation and the bias.  To do this 
compactly, we adapt the ``area figure of merit'' 
to the full risk.  Here, the dark energy equation of state 
$w(a)=w_0+w_a\,(1-a)$, where $a=1/(1+z)$ is the cosmic expansion factor, 
and the area of some likelihood contour in the $w_0$-$w_a$ plane 
is taken as the area figure of merit.  In practice, one equivalently 
quotes $1/[\sigma(w_a)\times\sigma(w_p)]$, where $w_p$ is the 
pivot value, the value of $w$ at the redshift where the uncertainties 
in $w_0$ and $w_a$ are uncorrelated.  To incorporate parameter 
biases $\delta p$, we define the risk figure of merit\footnote{In 
many circumstances 
this is a conservative estimate of the damage.  One could define an area 
taking into account all possible shifts of the likelihood contour due 
to bias, as effectively adding to the uncertainty.  This area increase 
is often larger than the effective area increase from the risk, but it 
is dependent on the $\Delta\chi^2$ level of the confidence contour 
considered, and so we stay with the well defined risk statistic.} 
from Eq.~(\ref{eq:risk}) as $1/[{\rm Risk}(w_a)\times{\rm Risk}(w_p)]$. 

\begin{figure}[!htb]
\begin{center}
\psfig{file=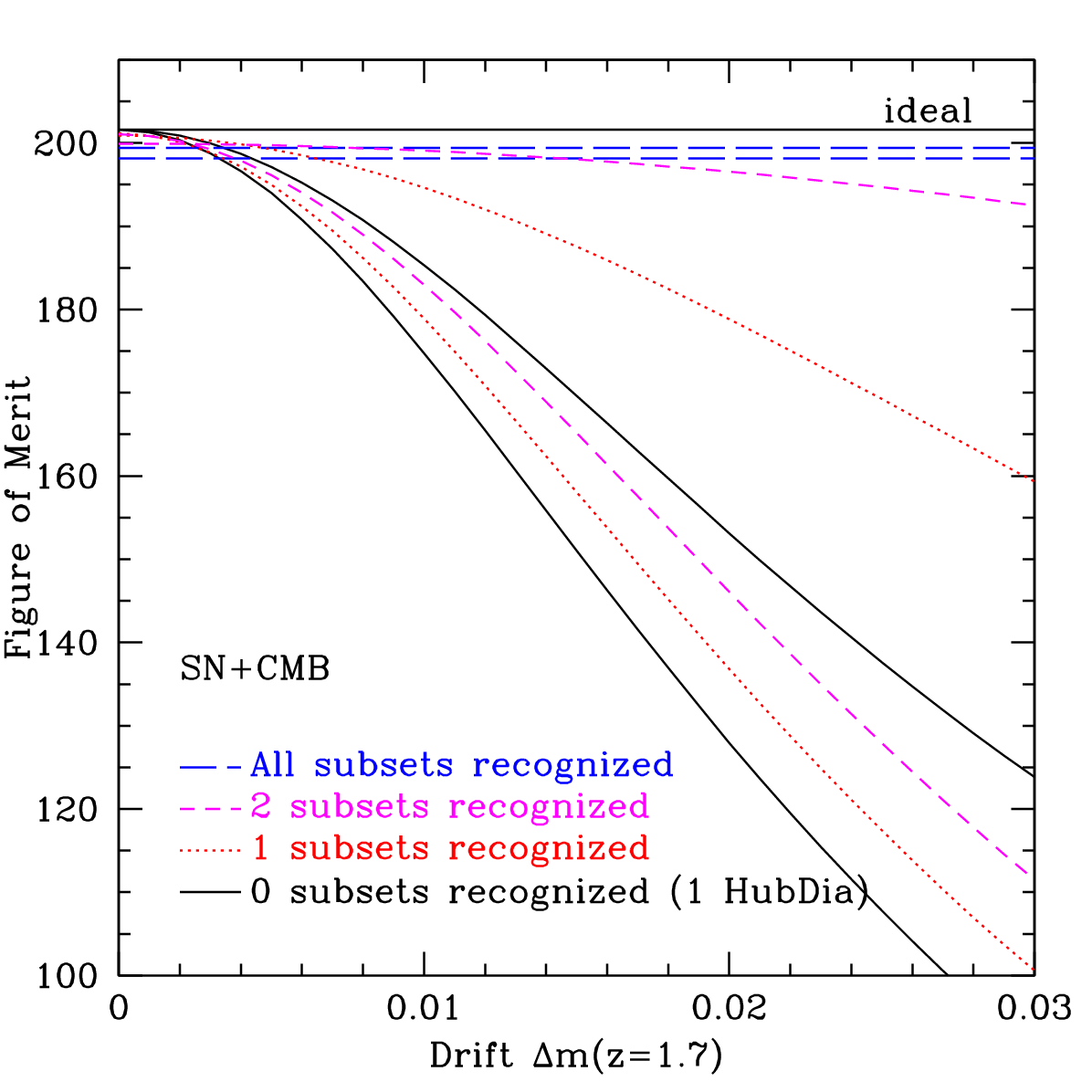,width=3.4in}
\caption{Recognition of like SN subsets has significant impact on the 
dark energy figure of merit incorporating the trade off between 
precision and bias.  
Unrecognized population drift induces evolution in the SN magnitude, 
$\Delta m(z)$, and bias in the cosmological parameters, while 
adding a fit parameter for recognized subsets costs in precision. 
For a single (full sample) Hubble diagram, the degradation in figure 
of merit due to bias can be substantial, as shown by lowest solid 
curve.  For each 
case we plot the envelope of worst and best results scanning over the 
population evolution and permutation of subsets recognized.  Maximizing 
the number of subsets recognized is the optimum strategy except for 
very small drifts, and even then the cost is less than 2\%. 
}
\label{fig:riskfom}
\end{center}
\end{figure}

Now we can quantify to what extent it is advantageous or not to 
rigorously define subsets through detailed observations.  
Figure~\ref{fig:riskfom} illustrates the effect on the dark energy 
parameter determination.  This combines simulated high quality data 
from 2300 SN between $z=0-1.7$ with Planck CMB information to estimate the 
cosmological parameters.  If we somehow knew that all subsets had the 
same mean absolute magnitude, i.e.\ that no magnitude evolution were 
possible, then the figure of merit is simply the usual area of merit 
and is shown by the horizontal line labeled ``ideal''.  If we use only 
a single Hubble diagram, making no effort to, or failing to, recognize 
subsets, then the degradation in figure of merit is severe, shown by 
the solid, black curves.  These show the best and worst cases of the 
values of $B$ used in the population drift.  For a total effective 
evolution to $z=1.7$ of 0.02 mag, the single Hubble diagram approach 
degrades the cosmology constraint by a factor of 1.3-1.6.  
No increase in the number of SN 
can fully make up for this degradation, assuming a systematic floor 
of $dm_{\rm sys}= 0.02(1+z)/2.7$.  Even worse, while larger numbers of 
SN will tighten the precision they will increase the relative bias on 
the cosmological parameters.

Recognizing 1 (dotted, red curves) or 2 (short dashed, magenta curves) 
of the 4 subsets 
acts to improve the situation.  (The case of 3 subsets recognized is 
equivalent to that of all recognized, since the remainder of the sample 
is simply the fourth subset.)  Here the upper and lower curves represent 
the best and worst of not only variation over $B$, but also the 
permutations of which of the 4 subsets are recognized.  That is, 
identifying the subset with the most extreme magnitude offset is most 
useful, while one with an offset little different from the mean is 
of marginal effect.  Indeed, if we recognize the two most extreme 
subsets, we approach the perfect situation, while finding the two least 
extreme ones only improves by 14\% over the worst case of the single 
Hubble diagram (also see \S\ref{sec:separated}). 

\begin{figure}[!htb]
\begin{center}
\psfig{file=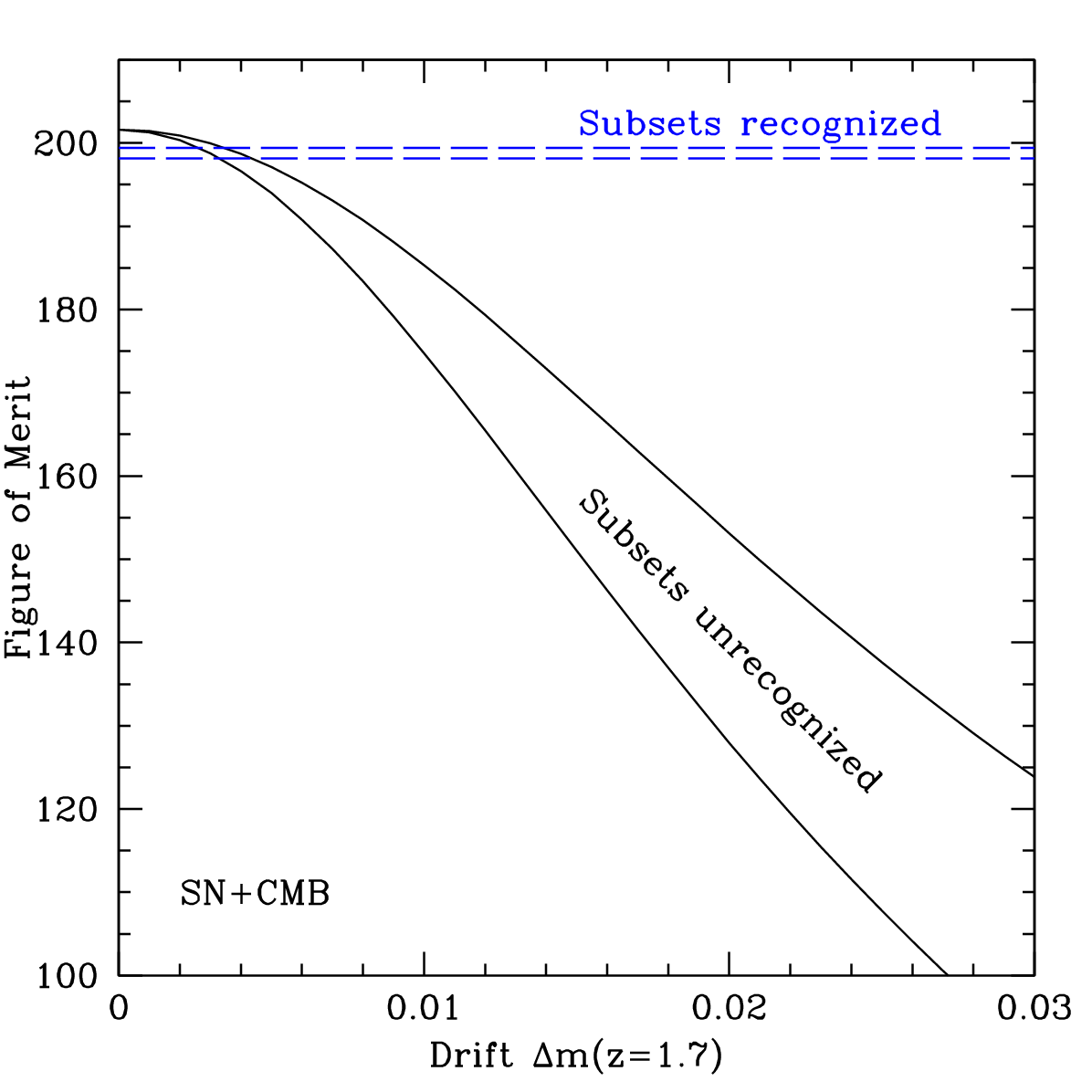,width=3.4in}
\caption{Same as Fig.~\ref{fig:riskfom} but a simpler version 
containing only the extreme cases to bring out the essential result. 
}
\label{fig:riskfomsimple}
\end{center}
\end{figure}

Finally, we consider sufficiently good observations to recognize all 
subsets (long dashed, blue curves).  In this case we must fit for 4 
different $\mathcal{M}$ 
parameters, and the key question was whether the elimination of bias 
was worth the loss in precision due to the expanded parameter set.  
The answer is emphatically yes -- the figure of merit is only 1.7\% 
below the ideal case.   This represents up to a 55\% improvement over 
using exactly the same SN in a single Hubble diagram (see 
Fig.~\ref{fig:riskfomsimple} for a clear view of these essential 
points).  Moreover, the 
answer obtained represents the true cosmology without a bias.  Only 
when the evolution is extremely small, $\Delta m(z=1.7)<0.005$, which 
we do not know a priori, do we fail to gain by employing the 
likes vs.\ likes approach, where again the highest cost is less than 2\%.

\section{Observational Requirements on Defining Subclasses \label{sec:obs}} 

While the optimum survey design would be to obtain a full suite of 
observations that 
enables recognition of all subclasses, this cannot always be realized. 
In this section we examine three cases of less than perfect observations 
and investigate the implications for cosmology determination.  Finally, 
while the issue of exactly how to define subclasses is complex and 
largely unknown, we discuss generically some possible routes toward this. 

The ability to recognize subsets depends on the acuity of the 
observations.  This in turn depends on  the instrumentation, exposure 
time, types of data collected, etc.  While this is too complex an area 
to explore here, we can get an idea of the effect on cosmology through 
toy models in the next three subsections, exploring respectively the 
degree of difference between subsets, overlap and confusion, and a 
continuum of subclass properties.

\subsection{Separated Subclasses \label{sec:separated}} 

Given disjoint subsets, with absolute magnitudes offset 
from the mean, one is most likely to recognize those subsets that are 
most discrepant.  Since these also tend to induce the greatest cosmological 
parameter bias (depending on the evolution of the subset fraction $f_i(z)$), 
this can mean that recognizing merely the most extreme subsets helps 
substantially toward removing bias. 

For example, in the results of Fig.~\ref{fig:riskfom} we find that 
recognizing the two most discrepant subsets gives a 28-52\% improvement 
over recognizing none (for a drift of 0.02 mag out to $z=1.7$), while 
recognizing the two least discrepant only improves by 8-14\%.   The 
absolute level from recognizing the two most discrepant subsets approaches 
2.5\% below the ideal case.  For recognizing a single subset, the 
improvements are 17-29\%  and 4-7\% for 
the most and least discrepant, respectively, and recognizing the most 
discrepant subset brings the figure of merit within 11\% of the ideal 
case. 

To examine this further, we consider the effect of increasing the 
ability to resolve the subsets.  For example, if we believe that 
some observable such as line velocity correlates with luminosity, 
then we need to have the capability to make sufficiently accurate 
measurements of this variable.  As a toy model we take subsets with 
absolute magnitudes distributed at $X$, $-(3/4)X$, $-X/2$, and 
$X/4$ relative to the mean at $z=0$ (recall from \S\ref{sec:cos} the 
sum of the $\Delta\mathcal{M}$'s is taken to be zero).  We then 
consider experiments with varying ability to resolve discrepancies 
from the mean and ask at what degree of difference does the figure of 
merit degrade by a certain percent. 

\begin{figure}[!htb]
\begin{center}
\psfig{file=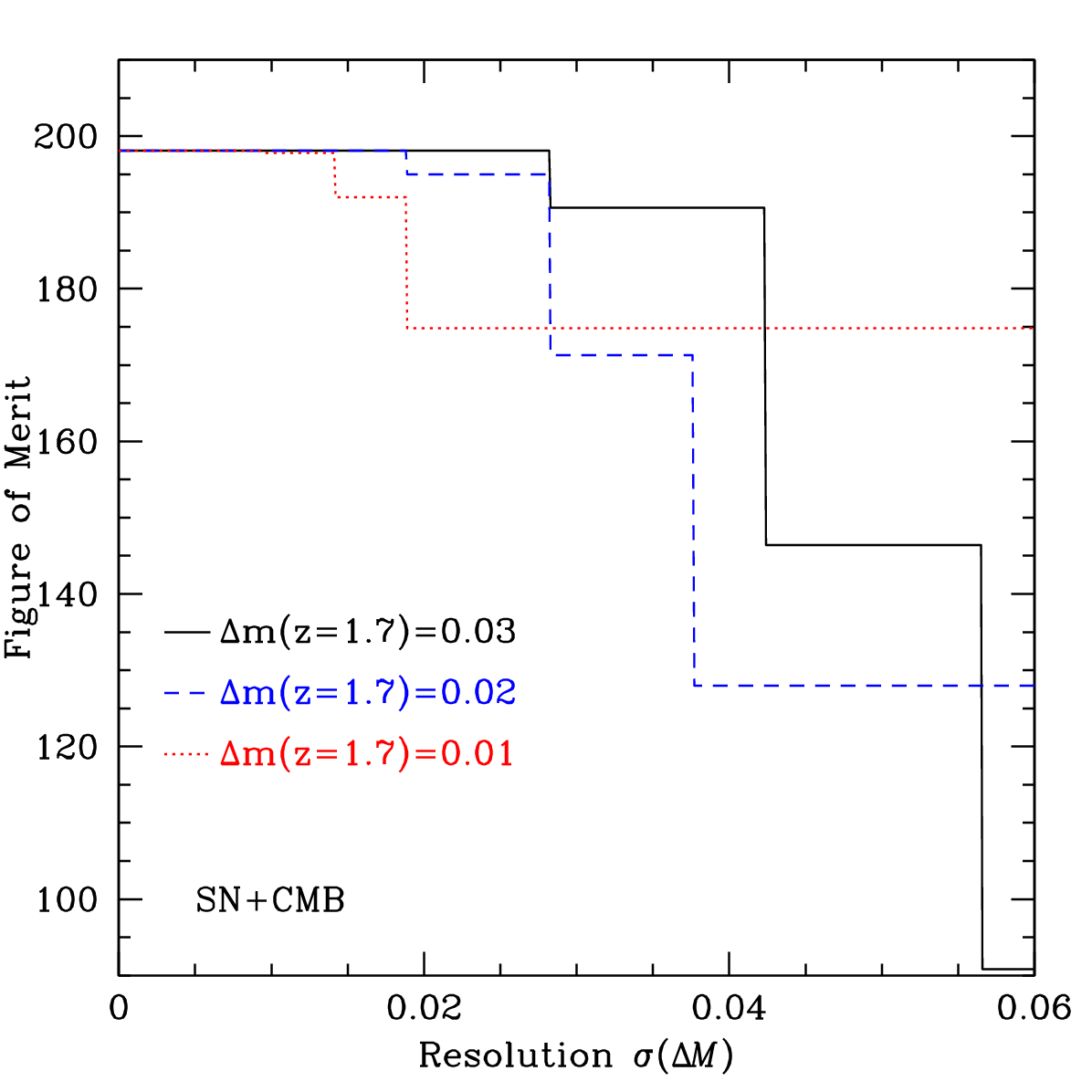,width=3.4in}
\caption{Finite resolution in measuring supernova characteristics 
translates into a finite resolution for distinguishing subsets from 
the mean sample luminosity.  For three levels of magnitude evolution 
we plot the degradation in dark energy figure of merit as resolution 
coarsens and subsets become unrecognized.  A rough rule of thumb is 
that an observational resolution less than the total evolution 
sensitivity to be probed, $\sigma(\Delta{\mathcal M})\lesssim 
\Delta m(z=1.7)$, is needed. 
}
\label{fig:resolve}
\end{center}
\end{figure}

As the resolution degrades past the smallest degree of difference 
of a subset from the mean, that subset is no longer recognized per se, 
but can still be identified as the ``leftover'' from all the other 
subsets.  Once the next subset threshold is passed, however, then 
the cosmology determination degrades, and so on as the resolution 
coarsens, until no subsets can be recognized.  Figure~\ref{fig:resolve} 
shows the behavior for the case of four subsets as before, though with 
the absolute magnitude distribution as above. 

The resolution required for no more than 10\% degradation in dark energy 
figure of merit is at the level of  0.019, 0.028, 0.042 mag for total 
drifts $\Delta m(z=1.7)=0.01$, 0.02, 0.03 respectively.  To limit the 
degradation to 20\% one requires resolution of 0.037, 0.042 
for $\Delta m(z=1.7)=0.02$, 0.03 (for only a 0.01 evolution in 
magnitude, the bias does not become large enough to reduce the figure 
of merit by 20\%).  In fact, these numbers are too optimistic in that 
one would need to see the subset deviate from the mean at 2-3$\sigma$ 
for robust recognition and suppression of normal outliers. 
In general, if the survey aims to be sensitive 
to an evolution at some level $\Delta m(z=1.7)$ (with the associated 
cosmological bias and leverage), then the observational resolution 
should be designed to be somewhat finer, $\sigma(\bar M)\approx 
(1/2)\,\Delta m(z=1.7)$.  

Of course when the experimental resolution weakens 
and makes discrimination of subsets from the mean difficult, the 
uncertainty effectively broadens the subsets and could cause them to 
overlap.  This is a situation distinct from straightforward recognition 
or not, and we discuss it next.

\subsection{Overlapping Subclasses \label{sec:overlap}} 

So far we have discussed the subsets as either recognized or 
unrecognized, and assumed that the recognized subsets are distinct. 
Subsets however can have some luminosity function distribution, as 
mentioned in \S\ref{sec:pdf}, and the recognition can be fuzzy.  For 
example, two subsets may possess overlapping luminosity functions and 
a member of one subset might be misassigned to another.  This will 
change the fraction in each subset away from the true value, inducing 
a cosmology bias 
\beq 
\delta m(z)=\sum_{\rm recognized} \Delta{\mathcal{M}}_i\,[\delta f_i(z) 
-\delta f_i(0)], 
\eeq 
where $\delta f_i$ is the misestimated population fraction.  Note that 
here the sum runs over recognized subsets, unlike in Eq.~(\ref{eq:unrecog});  
if the 
subsets are not recognized to begin with, then the mixing has no effect. 
Here $\Delta{\mathcal{M}}_i$ represents the true (though unknown) mean 
absolute magnitude of each subset, which we take to be unaffected by 
the misassignment (we consider fuzziness in both subset population {\it 
and\/} absolute magnitude in \S\ref{sec:continuous}).  

We examine the consequences for a model with a fraction $f_{12}$ of 
the total sample, belonging to subset 1 but overlapping with subset 2, 
having a probability $f_{1\to2}$ of these being misassigned.  In general, 
the misestimation gives 
\beq 
\delta f_i=\sum_{j\ne i} (f_{ji}P_{j\to i}-f_{ij}P_{i\to j}). 
\eeq 
Summing over all subsets (including the unrecognized ones) enforces 
that $\sum \delta f_i=0$, i.e.\ a supernova lost from one subset shows 
up in another, or in the main undifferentiated group.  

First consider the case of two overlapping subsets, and two extremes.  
If the misassignment leads to a complete swap of one subset with another, 
so that $f_{ji}=f_j$ and $P_{j\to i}=1$, then $dm(z)=X\,(A_2-A_1))(z/1.7)^B$ 
where the absolute magnitudes of the two subsets differ by $X$.  For the 
parameters of \S\ref{sec:risk}, this amounts to $dm(z=1.7)=-0.00125$, which 
is insignificant.  As the other extreme, if instead of a swap, a transfer 
occurs, i.e.\ a one-sided loss, with $P_{1\to2}=1$, $P_{2\to1}=0$, 
$f_{12}=f_1$ then $dm(z=1.7)=-X\,A_1=-0.0025$.  

These bias effects from 
recognized, but overlapping and confused, subsets will add to the bias 
due to the unrecognized subsets.  The overlap contribution is small, 
however, because one is not mistaking the absolute magnitude by the 
full deviation from the sample mean but only by the amount to the nearest 
subset's magnitude; furthermore, the biases from each subset are not 
additive but are differenced during an exchange.  Because the confusion 
is due to intrinsic luminosity function width then the observational 
resolution does not play a major role as in the previous section.  Indeed, 
with fine resolution one might be tempted to subdivide the sample into 
more subsets, which could lead to more overlaps, but such multiplicity 
of subsets further reduces the fractions $A_i$ and so the overlap biases 
are even smaller.

\subsection{Continuous Subclasses \label{sec:continuous}} 

Many supernova properties are not discrete, but continuous, and the 
subset categorizations may not be well separated, as discussed in the 
last two subsections.  
The limit of fuzziness in supernova properties is a continuous subclass 
distribution.  Here the sample is a cloud in some multidimensional 
observational data space and the absolute magnitude is a function 
of the location in that space.  We assume this is deterministic, so 
improved knowledge of the properties leads to a tighter distribution 
for the absolute magnitude.  This means that bias due to unrecognized 
subsets is replaced by bias due to unpinpointed properties.  In 
the completely unlocalized case (where observations are too weak to 
determine the location within the cloud, e.g.\ missing some type of 
observation can make the uncertainty in some dimension span the entire 
range), this is equivalent to the case of no recognized subsets, i.e.\ 
a single Hubble diagram. 

We can adapt the formalism of the discrete subsets by taking the sum 
over subsets to an integral over continuous variables.  If 
$\vec\pi$ represents the multidimensional parameter set over properties 
$x_1$,\dots $x_N$ (e.g.\ metallicity, velocity decline rate, silicon 
line ratio, etc.), then 
\beq 
\Delta m(z)=\int d\vec\pi\,\Delta{\mathcal M}(\vec\pi)\,[f(\vec\pi,z)- 
f(\vec\pi,0)]. \label{eq:dmcont}
\eeq 

For simplicity, we first consider a one dimensional space over a continuous 
property parametrized by $x$.  Both $\Delta{\mathcal M}$ and $f$ will 
be functions of $x$.  Taking 
\beq 
f(x,z)-f(x,0)=\Delta F(x)\,(z/1.7)^B\,, \label{eq:fF}
\eeq 
the mean value of the parameter $x$ drifts from $x_0=\int dx\,xf(x,0)/
\int dx\,f(x,0)$ to 
\beq 
\langle x\rangle(z)=x_0+(\frac{z}{1.7})^B\int dx\,x\,\Delta F(x). 
\eeq 
This drift causes a change in the mean absolute magnitude of the full 
sample (recall there are no individual subsets in this approach), 
$\Delta{\mathcal M}(\langle x\rangle(z))$.  

To evaluate Eq.~(\ref{eq:dmcont}) further, we must adopt forms for 
$\Delta F(x)$ and $\Delta{\mathcal M}(x)$.  Suppose 
\beqa 
\Delta F(x)&=&\Delta_F\,(x^n-x_F^n) \label{eq:dfrac} \\ 
\Delta {\mathcal M}(x)&=&\Delta_M\,(x^p-x_M^p)\,, \label{eq:dmag} 
\eeqa 
so the values $x_F$, $x_M$ define the standards.  For example, if 
$x$ represents metallicity, then a supernova with $x=x_M$ defines 
the baseline in absolute magnitude, and supernovae with $x=x_F$ maintain 
a constant population fraction, i.e.\ do not drift (the value $x_F$ 
does not have to actually be realized in the sample).  As the value 
of $x$ deviates from the 
standards, the demographics changes according to Eq.~(\ref{eq:dfrac}), 
with lower metallicity supernovae becoming more common at high redshift, 
say, and their luminosity changing according to Eq.~(\ref{eq:dmag}), with 
lower metallicity supernovae being brighter, say.   

The completeness conditions are $\int dx\,\Delta F(x)=\int dx\, 
\Delta{\mathcal M}(x)=0$, i.e.\ every supernova lies somewhere 
in the parameter space and there is a mean absolute magnitude for 
the sample.  Then over some finite range $x\in[x_-,x_+]$, 
\beq 
x_F=\left[\frac{x_+^{n+1}-x_-^{n+1}}{x_+-x_-}\frac{1}{n+1}\right]^{1/n}, 
\eeq 
and the equivalent for $x_M$ with $p$ substituting for $n$. 

The magnitude offset generating bias in the cosmological parameter 
estimation then takes the form for the continuous case 
\beq 
\Delta m(z)=\Delta_F \Delta_M\, (z/1.7)^B \int dx\,(x^n-x_F^n)(x^p-x_M^p)\,. 
\eeq 
For $n=p=1$ this gives 
\beq 
\Delta m(z)=(1/12)\Delta_F \Delta_M\,(x_+-x_-)^2 (z/1.7)^B\,. 
\eeq 
We recognize the maximum drift for the sample is $\Delta F_{\rm max}=
\Delta_F\, (x_+-x_-)$, with a similar expression for $\Delta {\mathcal M}$, 
so $\Delta m(z)=(1/12)\Delta F_{\rm max} \Delta M_{\rm max} (z/1.7)^B$.  
An analogous expression holds for other values of $n$, $p$.  

To generalize to a multidimensional parameter space over $x_1,\dots,x_N$, 
we have 
\beqa 
\Delta m(z)&\!\!=\!\!&(z/1.7)^B V_\pi^{-1} \sum_i \Delta_{F,i}\Delta_{M,i} 
\times \nonumber \\ 
&\!\!\qquad\!\!{}&\int dx_i\,(x_i^{n_i}-x_{F,i}^{n_i})(x_i^{p_i}-
x_{M,i}^{p_i}), 
\eeqa 
where $V_\pi=\int dx_1\dots dx_N$ is the parameter space volume and we 
have assumed the parameters $x_i$ are independent of each other.  

So far we have considered that we can measure the observational parameters 
$x$ with perfect accuracy and with this possibly determine the 
demographics $f(x)$.  Of course {\it if\/} knowing $x$ allows us to predict 
$\Delta{\mathcal M}(x)$ as well, then we can compute $\Delta m(z)$ and 
there will be no cosmology bias.  But now we consider the case where 
the measurements are not perfect but have some uncertainty $\delta x$, 
which will propagate through the demographics and absolute magnitude 
into the magnitude offset.  This is similar to the ``fuzzy'' philosophy 
of \S\ref{sec:overlap}. 

The measurement imprecision will lead to an magnitude uncertainty 
\beqa 
\delta m(z)=\int dx\, \delta x(x) &\!\!\Biggl\{\!\!& 
\frac{\partial\Delta{\mathcal M}(x)}{\partial x}\, 
[f(x,z)-f(x,0)] \label{eq:sigmam} \\ 
&\!\!{}+\!\!&\Delta {\mathcal M}(x)\left[\frac{\partial f(x,z)}{\partial x}- 
\frac{\partial f(x,0)}{\partial x}\right]\Biggr\}. \nonumber 
\eeqa 
This can be viewed as taking place in a multidimensional parameter space 
of $\vec x$ as well.  If there is no uncertainty in some $x_i$, or if $f$ 
and $\mathcal{M}$ are independent of $x_i$ then this dimension does not 
contribute to the magnitude systematic.  For simplicity we will write the 
expressions in terms of a single continuous parameter $x$.  

Using the forms of Eqs.~(\ref{eq:fF}), (\ref{eq:dfrac}), (\ref{eq:dmag}), 
we can evaluate the magnitude systematic in Eq.~(\ref{eq:sigmam}) given 
some observational input for the uncertainty $\delta x(x)$.  As the 
simplest case, we take $\delta x$ constant.   For $n=p=1$ the 
completeness conditions ensure that the systematic is zero.  The general 
result is of the form 
\beq 
\delta m(z)\approx \Delta{\mathcal M}_{\rm max} \Delta F_{\rm max} 
\frac{\delta x}{\Delta x}\,\left(\frac{z}{1.7}\right)^B. \label{eq:dmf} 
\eeq 
The range $\Delta x$ can be defined through either theoretical model 
limits on the variation of $x$ (e.g.\ metallicity) or as some 
weighted range that captures 90\%, say, of the magnitude drift.  

Thus, perfect measurements give no uncertainty in this situation where 
the functional dependences are assumed known, but as the observations 
become more imprecise, i.e.\ $\delta x$ increases, the magnitude 
uncertainty grows.  Eventually 
the perturbative formalism used here breaks down, but when $\delta x$ 
becomes comparable to $\Delta x$ then this approach should reduce to 
the single Hubble diagram case. 

\begin{figure}[!htb]
\begin{center}
\psfig{file=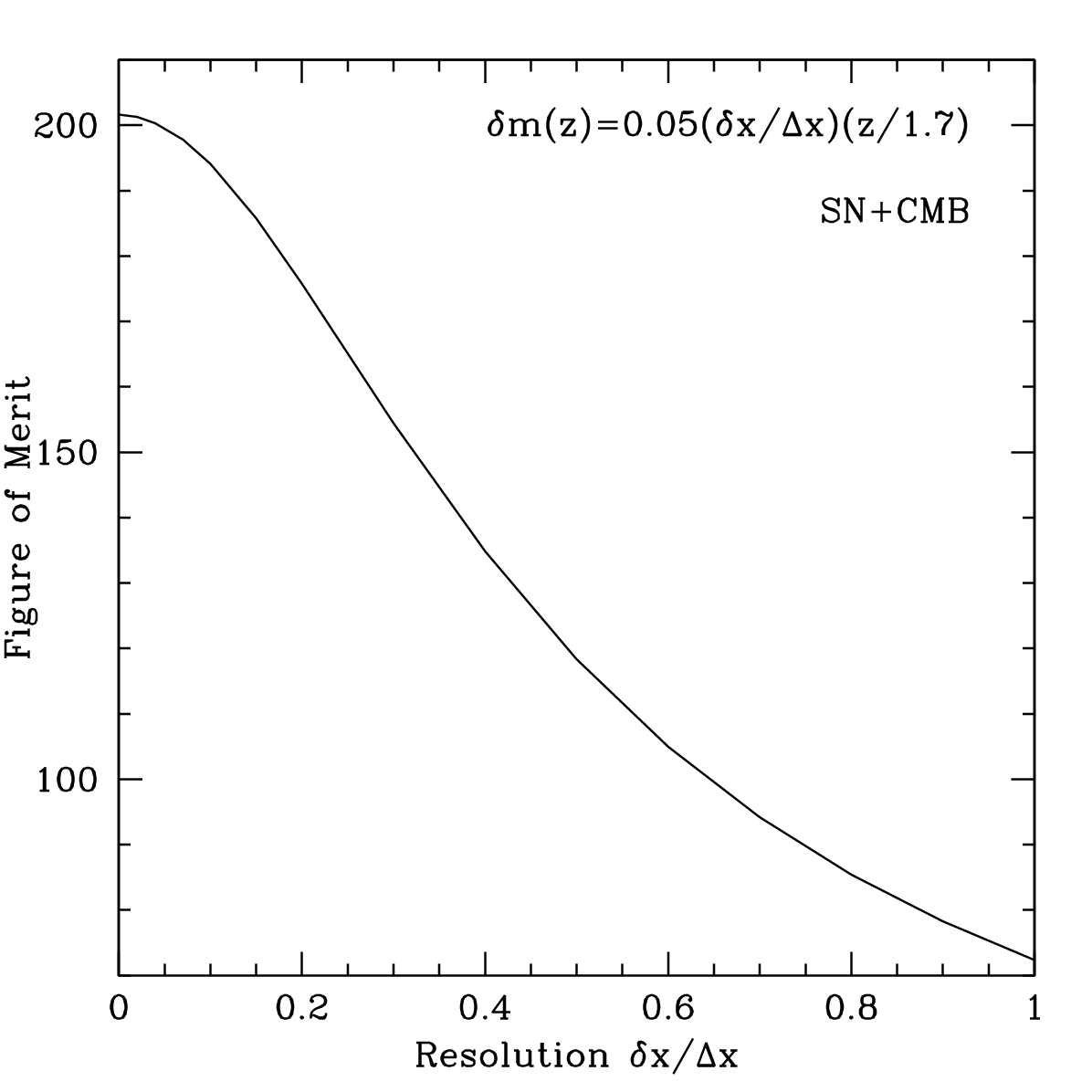,width=3.4in}
\caption{For continuous parameters defining supernova subclasses, 
lack of observational resolution degrades the dark energy figure of 
merit.  When the resolution $\delta x/\Delta x=0$, then the observations 
exactly determine the supernovae properties.  When $\delta x/\Delta x=1$ 
then the observations are blurred over the entire sample, making this 
equivalent to using only a single Hubble diagram, but with an added 
fit parameter for the drift amplitude, reducing the figure of merit by 
a factor of $\sim3$. 
}
\label{fig:contres}
\end{center}
\end{figure}

One can remove the bias due to the lack of observational resolution by 
fitting for the form of the magnitude offset, e.g.\ Eq.~(\ref{eq:dmf}). 
Two fit parameters are the evolution power index $B$ and the prefactor,  
call it $C$.  If no prior is placed on these quantities, 
then the degradation on dark energy parameters is severe, reducing the 
figure of merit to less than 2.  In particular, $B$ is poorly determined 
and covariant with the dark energy variables, so that even an overidealized 
prior of 0.002 on $C$ gives a figure of merit of only 62.  We therefore 
fix $B=1$ and investigate the degradation as a function of the prior 
on $C$, essentially equivalent to observational resolution $\delta x/ 
\Delta x$.  Figure~\ref{fig:contres} shows the results as a function 
of this resolution.

The figure of merit is rapidly degraded as the observational acuity 
decreases.  The effective total magnitude offset here is 
$\delta m(z=1.7)=0.05\, 
(\delta x/\Delta x)$, so a resolution of 0.4 corresponds to 0.02 mag 
evolution.  To defend against degradation of more than 20\% in the 
figure of merit requires a resolution of 0.27.  Of course as more 
fit parameters are added, the requirements will tighten.  Thus, lack 
of observational resolution leads directly to unpinpointed or confused 
subclasses and loss of cosmological information.

\subsection{Defining Subclasses \label{sec:defsub}} 

A central issue mentioned in \S\ref{sec:pdf} is the consequence when 
a subclass fails, i.e.\ when a carefully 
characterized subset does not have an unevolving mean luminosity. 
(Recall we don't require the luminosity distribution to be independent 
of redshift, only that the mean stays constant.)  If the subset is 
not a true subclass then we can absorb the residual luminosity 
evolution into an effective population drift $\tilde f_i(z)$ via 
\beq 
f_i(z)\,L_i(z)=f_i(z)\,\frac{L_i(z)}{L_i(0)}\,L_i(0) 
=\tilde f_i(z)\,L_i(0)\,. 
\eeq 
So a drift in $L_i(z)$ because the subset $i$ is not a true 
subclass can be viewed as an uncertainty in $\tilde f_i(z)$.  We can then 
try to account for this by fitting for $\tilde f_i(z)$.  
(Note that now the quantity $\tilde f_i(z)$ is not directly observable.)  
As a fitting function 
we consider an expansion in Chebyshev polynomials over the range 
$z=0-1.7$.  We include terms through second order so as to allow the 
possibility for non-monotonic behavior, with 
\beq 
\tilde f_i(z)=\sum_{j=0}^2 \alpha^{(i)}_j\, T_j(x=z/1.7), 
\eeq 
where we normalize the polynomials to the interval $[0,1]$.  

\begin{figure}[!htb]
\begin{center}
\psfig{file=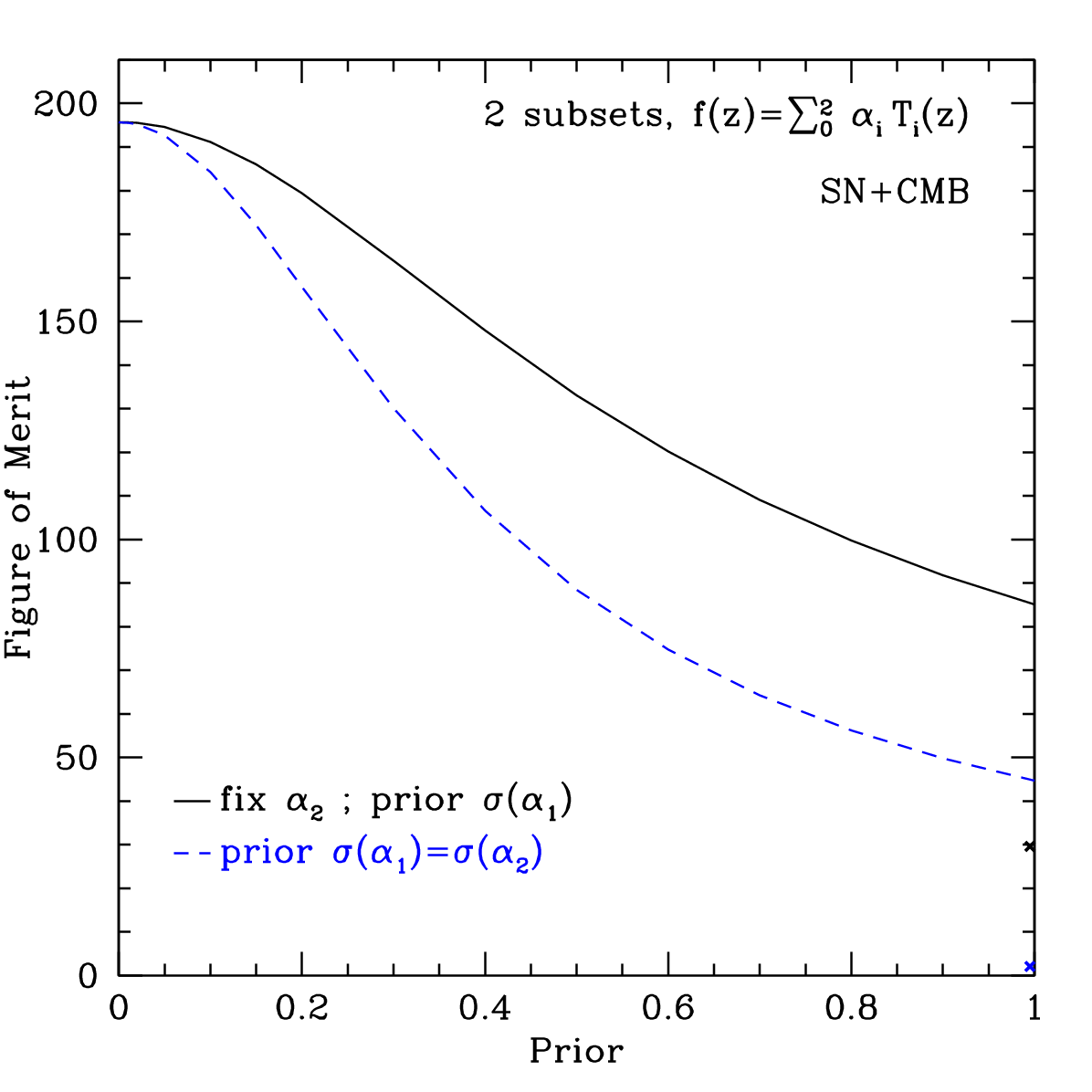,width=3.4in}
\caption{Unrecognized magnitude evolution can be accounted for by fitting 
for an effective population drift $f(z)$.  Here the evolution is expanded to 
second order in Chebyshev polynomials and the fiducial parameters 
correspond to $\Delta m(z=1.7)=0.02$ mag.  We show the dark energy 
figure of merit as a function of the priors placed on the subset 
evolution expansion coefficients.  The solid, black curve corresponds to 
using only a first order Chebyshev expansion while the dashed, blue 
curve uses a second order expansion.  In the first case the prior 
refers to that on $\alpha_1$ while in the second case we take the 
priors on $\alpha_1$ and $\alpha_2$ to be equal.  When no priors are 
used, the value of the figure of merit is shown by the x's at the 
right axis.  The magnitude of the priors can be related to an additional 
evolution by $\delta m_{\rm max}=0.02\,[\sigma(\alpha)/0.5]$ for 
either coefficient $\alpha_i$. 
}
\label{fig:tcheb}
\end{center}
\end{figure}

Adding such freedom degrades the figure of merit unless tight priors 
are placed on the amplitude of magnitude evolution allowed within 
the subset.  Figure~\ref{fig:tcheb} explores the effect of fitting 
for a residual magnitude evolution of amplitude $\Delta m(z=1.7)=0.02$, 
considering two subsets that fiducially linearly evolve from equal 
fractions at $z=0$ to 100\% in one subset at $z=1.7$ (i.e.\ the 
fiducial case is $\alpha_0=0.5$, $\alpha_1=0.5$, $\alpha_2=0$).  
We have further simplified the situation by taking 
$\tilde f_2(z)=1-\tilde f_1(z)$, which will not be true in general 
since $\tilde f_i$ no longer represent physical fractions of the sample. 
If this were relaxed or more subsets were used, the number of fit 
parameters increases and the degradation worsens.

Without priors on either Chebyshev coefficient, $\alpha_1$ or $\alpha_2$, 
the figure of merit plunges by a factor 100 to a value of 2.  Even 
freely fitting one coefficient 
lowers the figure of merit to 30.  Priors on each $\alpha$ of 0.5, 
corresponding to a maximum evolution uncertainty of 0.02 mag from 
each, degrades the figure of merit by a factor 2.2 (1.5 if only allowing 
linear evolution). 

The size of the effect due to residual uncertainty in whether the subset 
is truly a subclass points up the importance of having a comprehensive 
suite of precision observations.  Exactly what these should be is not yet 
known. The supernova spectrum should contain 
the required information (see for example 
\cite{bronder,foley,garavini,branch06,benetti}); 
broad band photometry may not be 
sufficient.  Recall that $\sim$60\% of the bolometric flux is emitted in 
the rest frame $BVR$ bands, so relying only upon rest frame 
ultraviolet or near infrared measurements leaves open the possibility 
that the tail does not move in the same way the dog does.  Similarly, 
another area of active research involves the use of particular spectral 
features \cite{nugent95,hatano00,blondin,ellis08}.  While any of these 
may prove robust, global analysis of the 
supernova spectrum appears less subject to such uncertainties.  One 
way to implement this could be through principal component analysis 
(PCA) for example (see \cite{tampca,naopca} for early steps).  

PCA could effectively tell us whether the defining subclass variables 
involve, e.g., line ratios, velocities, velocity changes, etc.  
While the amount 
of degradation was significant when adding only two extra fitting 
parameters in the Chebyshev polynomial case, PCA by its nature focuses 
on the most relevant combinations of variations, and so may prove a 
tractable analysis approach in combination with spectral observations.  
Indeed, preliminary indications point to the first two PCs accounting 
for 85\% of the spectral variation \cite{naopca}.

\section{Conclusions \label{sec:concl}} 

Without any systematics, Type Ia supernovae would be statistically the 
most powerful tool for probing the accelerating expansion of the universe. 
One of the key approaches for controlling systematics is that of likes 
vs.\ likes, or supernova demographics, carefully comparing sample 
properties through a suite of observational characterizations.  The simple 
concept is that like supernovae at different redshifts 
accurately reveal the cosmology, while supernovae at the same redshift, 
differing in essential ways, can define subsets giving clues to reining 
in systematics.  

The issue is not one of evolution, but {\it uncorrected\/} evolution 
and unrecognized evolution.  This article examines techniques for evaluating 
the cosmological consequences of systematic control or the lack of it, 
and strategies for implementing such control.  The main pitfall is bias 
of the cosmological parameters -- this is a bad thing, not just because 
it degrades the effective dark energy figure of merit calculated in terms 
of the risk, but because physics is lost.  One may end up with an 
impressively precise but simply inaccurate conclusion. 

Having high observational acuity and using all this 
information to define robust subsets is the optimal strategy.  We 
quantify this and demonstrate that this holds even at the price of 
additional subset parameters in the fit.  Analyzing 
the data in a single Hubble diagram can lead to biases of order a full 
statistical sigma and (secondarily) loss of figure of merit by 
a factor 1.6.  To avoid these consequences, one uses the recognized 
subsets to add fit parameters; this restores essentially all the 
cosmological leverage, as long as the subsets are sufficiently well 
defined by the observations that these subset mean luminosities do not 
evolve. 

The observational requirements to define the subsets is a complex 
subject but we consider three categories, of separated, overlapping, 
and continuous, or unrecognized, confused, and unpinpointed, 
subsets, and quantify some requirements within 
simplistic models.  We further briefly consider subsets whose 
luminosities do in fact evolve and speculate that principal component 
analysis applied to supernovae spectra may prove the best path to 
robust control.  In the appendix we illustrate how combining separate 
data sets, especially from different redshift ranges, can act similarly 
to evolution and have significant deleterious effects. 

A supernovae survey designed without the controls enabled by 
high observational acuity is taking a gamble on the astrophysics 
and supernova properties being kind.  While we do not yet know exactly 
what subsets to define, the capability and flexibility to do so, as 
measured quantitatively along the lines of the simple calculations 
here, are required to ensure confidence in the cosmological results.

\acknowledgments 

I thank Bob Cahn, Ariel Goobar, Dragan Huterer, Alex Kim, Peter Nugent, 
Reynald Pain, and Saul Perlmutter for useful discussions. 
This work has been supported in part by the Director, Office of Science, 
Office of High Energy Physics, of the U.S.\ Department of Energy under 
Contract No.\ DE-AC02-05CH11231.

\section*{Appendix: Redshift Distribution Effects \label{sec:box}} 

The redshift dependence of the population drift, or more generally 
the subset distribution, convolves with the Fisher sensitivity 
derivatives $\partial m/\partial p_j$ in Eq.~(\ref{eq:bias}) in a 
complicated manner to lead to parameter bias.  One cannot in general 
predict analytically how a given form for $f_k(z)$ leads to a bias. 
The offset $\Delta m(z)$ beats against $\partial m/\partial p_j$ but 
the set of $\partial m/\partial p_j$ do not form a complete basis, 
nor even an orthogonal one.  Even if $\Delta m(z)$ had exactly the 
same functional form as some $\partial m/\partial p_j$, the offset 
propagates not just to the parameter $p_j$ but to all the parameters 
(unless the 
inverse Fisher matrix in Eq.~\ref{eq:bias} is formed purely from 
SN magnitude information, without CMB information or priors).  The 
one exception is a redshift independent $\Delta m(z)$, which induces 
a pure shift in $\mathcal{M}$ since this parameter enters only into 
SN magnitudes. 

Thus we must calculate the effect of various forms of $f(z)$ 
numerically, and it is important to consider a variety of behaviors 
as we do.  In general, we find that nearly linear redshift evolution, 
$B\approx 1$, has the greatest impact on the risk figure of merit. 

However we could consider another type of magnitude offset, not intrinsic 
to the SN populations, but rather the measurement process.  If 
different surveys are combined, a miscalibration between the 
magnitudes can ensue, even if the SN absolute magnitudes are equal, 
due to filter or instrumental zeropoint offsets.  The redshift 
dependence of the samples, taking the place of population drift 
$f(z)$, can be particularly sharp, for example when combining a 
lower redshift ground-based sample with a high redshift space-based 
sample.  As one example, if these sets are matched 
at $z=0.8$ with an unrecognized miscalibration of 0.02 mag, then 
the cosmological parameter bias causes the risk figure of merit to 
be degraded by a factor 2.7 (with parameters biased by up to 
$1.9\sigma$).  See Fig.~17 of \cite{linrpp} for other 
matching scenarios.  Overlap between the sets needs to be substantial 
to ameliorate the degradation. 

In general one would want to define new fit parameters for 
possible offsets when using multiple samples, to eliminate bias, 
but these additional parameters tend to increase the dispersion 
substantially.  For example, with a single offset fit parameter 
the area figure of merit degrades by a factor 2.4 without a prior 
on the offset; the factor is still 1.6 with a prior of 0.02 mag. 
So to add to the other strategies for controlling systematics, 
a homogeneous sample over the full redshift range, or substantial 
overlap between sets, strongly improves the cosmological accuracy.

\end{document}